\documentclass[a4paper,11pt]{article}

\usepackage{amsmath,amsfonts}
\usepackage{bm}
\usepackage{ascmac}
\usepackage{fancybox}
\usepackage{amsthm}
\usepackage{amssymb}
\usepackage{physics}
\usepackage{ulem}
\usepackage{multicol}
\usepackage{multirow}
\usepackage{cite}
\usepackage{graphicx}
\usepackage{wrapfig}
\usepackage{hyperref}
\usepackage{caption}
\newcommand{\myfigure}[4][width=8cm]{
\begin{figure}[htbp]
\centering
\includegraphics[#1]{#2}
\caption{#3}\label{fig:#4}
\end{figure}}

\begin{document}
\title{Space versus Context: Competition for limited neural resources determines engram cell allocation in the hippocampus}
\date{}
\author{Kensuke Chiba, Jun-nosuke Teramae}

\maketitle
% \clearpage
% \tableofcontents
% \clearpage

\section{Abstract}
Engram cells in hippocampal CA1 represent contextual information associated with events experienced by animals and constitute a cellular substrate of episodic memory. Recent experiments have shown that a subset of hippocampal place cells is recruited as engram cells for individual contexts. However, the principle governing engram cell allocation across contexts remains unclear. Here, we develop an information-theoretic framework that quantifies the trade-off between spatial and contextual information encoded by CA1 neurons. Our theory shows that limited neural resources inevitably create competition among contexts for the recruitment of place cells as engram cells. Solving this optimization problem determines the fraction of engram cells allocated to each context as a function of its occurrence probability. The theory predicts a non-monotonic relationship between context probability and engram allocation, reflecting competition between spatial and contextual information under resource constraints. Consequently, the recruited fraction emerges through a discontinuous transition at a critical probability, reaches a maximum at intermediate probabilities, and decreases for highly frequent contexts. Furthermore, we analytically derive the phase boundary for the emergence of finite engram cell allocation as a function of contextual input strength and the relative importance of contextual information. These findings identify competition for limited neural resources as a key determinant of engram cell allocation and provide experimentally testable predictions for how memory representations are distributed across contexts under resource constraints.

\section{Introduction}

The concept of a physical substrate of memory, termed the engram, was first proposed in the early twentieth century. Although direct experimental identification of memory engrams remained elusive for decades, recent advances in experimental techniques, including activity-dependent genetic labeling and causal optogenetic manipulation of targeted neurons, have enabled the identification of engram cells that encode and store individual memory events \cite{loss_of_func, garner2012generation, gainfunc, ramirez2013creating}.

For each contextual event experienced by an animal, a sparse ensemble of neurons is recruited as its engram. These neurons are activated during memory formation and preferentially reactivated during memory retrieval (Fig.~\ref{fig:1} (a))\cite{josselyn2015finding, barron2017inhibitory, kitamura2017engrams, engram_PC, engram_review2, denardo2019temporal, frankland2019neurobiological, engram, goode2020integrated, ryan2022forgetting, choucry2024engram, tome2024dynamic, kobayashi2026neural, delamare2026computational}. It has also been reported that neurons with relatively high intrinsic excitability at the time of memory formation are more likely to be recruited into the engram\cite{han_engram_competition, zhou_engram_competition, yiu_engram_competition, delamare2024intrinsic, ghandour2025parallel}. However, the principle governing engram cell allocation across contexts remains unclear\cite{kastellakis2016linking, engram_allocation, gastaldi2021shared}. In particular, it remains largely unknown why only a sparse subset of neurons is recruited for each context and what determines the fraction of the neuronal population allocated to the engram of a given contextual event.

To address these questions, we focus on a recent experimental finding that engram cells in hippocampal CA1 are recruited from a subset of place cells\cite{engram_PC}, which encode the animal's spatial location \cite{treves1992computational, samsonovich1997path, zhang1998interpreting, barry2006boundary, solstad2006grid, moser2008place, romani2015short, stachenfeld2017hippocampus}. Importantly, when a place cell is recruited as an engram cell, its place field, the spatial region in which the neuron preferentially responds, is significantly expanded (Fig.~\ref{fig:1} (b))\cite{engram_PC}. Because the expansion of a place field reduces the spatial resolution of the neuronal response, this finding suggests that engram cell recruitment occurs at the cost of spatial information represented by CA1 neurons. Thus, engram cell allocation may reflect a trade-off between spatial information and contextual information. Since the number of neurons is finite in CA1, this trade-off may provide a principle for determining how limited neural resources are allocated between these two types of information.

Here, we formulate this trade-off using mutual information\cite{shannon1948mathematical, cover1991elements}. Mutual information provides a quantitative measure of statistical dependence between random variables and has been widely used to study neural coding and efficient information representation in the brain \cite{atick_efficient_coding, brunel1998mutual, deneve1999reading, averbeck2006neural, zhang2023hippocampal, tatsukawa2025cortical}. We derive analytical expressions for the mutual information between neuronal activity and variables describing the animal's spatial location and behavioral context. By solving the resulting optimization problem, we determine the fraction of place cells recruited as engram cells in CA1 as a function of the probability of the corresponding context, the strength of contextual input, and the relative weight assigned to contextual information.

The theory provides nontrivial yet biologically plausible predictions. The optimal fraction of engram cells undergoes a discontinuous transition from zero to a finite value when the corresponding context probability exceeds a critical threshold. Moreover, this finite optimal fraction is maximized at an intermediate context probability and decreases as the context probability becomes large. The theory further predicts two distinct phases in the optimal allocation of engram cells. In one phase, no engram cells are allocated to any context, whereas in the other, a finite fraction of place cells is recruited as engram cells. We analytically derive the phase boundary between these phases as a function of contextual input strength and the relative weight of contextual information. These results suggest that efficient information coding under limited neural resources provides a theoretical principle for engram cell allocation and offers experimentally testable predictions for the sparse recruitment of engram cells in hippocampal CA1.

\myfigure[width=12cm]{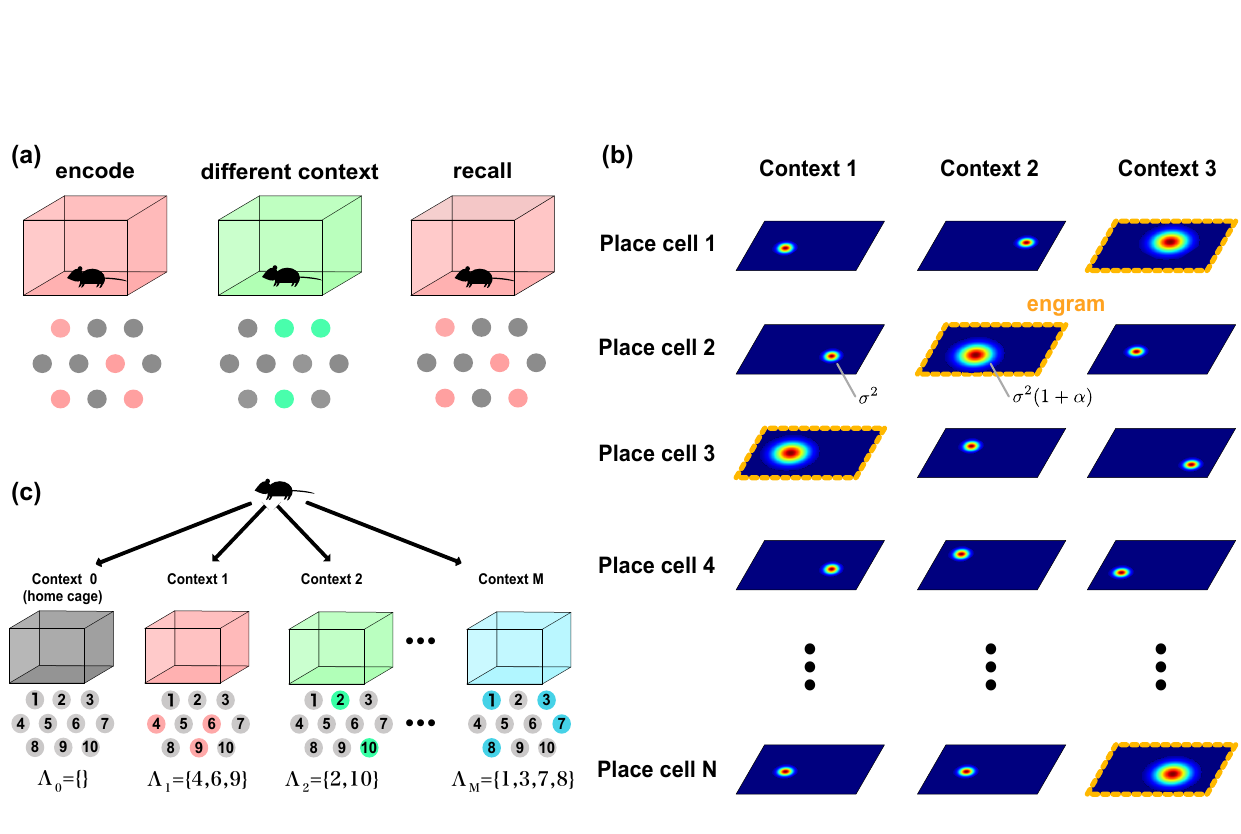}{Engram cells that store contextual events experienced by an animal are recruited from a subset of place cells, which encode the animal’s spatial location in hippocampal CA1. (a) When an animal experiences a novel environment, a subset of hippocampal CA1 neurons is selectively activated and recruited as engram cells for the context. Different subsets of neurons are recruited as engram cells for different contexts. These neurons are reactivated when the animal experiences the same environment and recalls the context. (b) When a place cell is recruited as an engram cell for a specific context, its place field expands in the corresponding context\cite{engram_PC}. (c) We model engram cell allocation as a process in which subsets of neurons, denoted by $\Lambda_s$, $s=0,\ldots,M$, are recruited as engram cells for the corresponding contexts. The animal experiences one of the $M+1$ contexts, including the home cage, with different occurrence probabilities.}{1}

\section{Method}

\subsection{Model}

We consider an animal exposed to $M+1$ contexts, denoted by $C_s$ $(s=0,1,\ldots,M)$, where $C_0$ represents the home cage (Fig.~\ref{fig:1} (c)). The animal is placed in context $C_s$ with probability $p_s$. For simplicity, each context is modeled as a two-dimensional environment with the same area $S$, and the animal is assumed to explore each environment uniformly. The animal's position within a context is represented by a two-dimensional vector $\boldsymbol{x}=(x,y)^\top$. Thus, the animal's state is specified by the pair $(s,\boldsymbol{x})$, and its probability density is given by $P(s,\boldsymbol{x})=p_s P(\boldsymbol{x}\mid s)=p_s P(\boldsymbol{x})=p_s/S$.
Note that $s$ and $\boldsymbol{x}$ are, therefore, statistically independent under this setting.

Let $N$ denote the number of CA1 place cells. For each context, a subset of these neurons is recruited as engram cells associated with that context. We denote by $\Lambda_s$ the set of indices of place cells recruited as engram cells for context $C_s$. For simplicity, we assume that the sets $\{\Lambda_s\} $ are mutually disjoint. We also assume that no engram cells are assigned to the home cage, so that $\Lambda_0=\emptyset$.

The place field of the $i$th CA1 place cell is described by a function $\lambda_i(\boldsymbol{x},s)$, which represents the firing rate of the neuron when the animal is at position $\boldsymbol{x}$ in context $C_s$, where $i=1,\ldots,N$. Experimental studies have shown that place fields expand when neurons are recruited as engram cells for the corresponding context\cite{engram_PC}. Accordingly, we model each place field as an isotropic two-dimensional Gaussian with its spatial extent determined by the neuronal identity $i$ and the context $s$:
\begin{align}
\label{equ:lambda_i}
    \lambda_i(\boldsymbol{x},s)
    &\propto \exp\left(
    -\frac{
        \left|\boldsymbol{x}-\boldsymbol{\mu}_i(s)\right|^2
    }{2\sigma_i(s)^2}
    \right),\\
    \sigma_i(s)^2
    &= \left\{
    \begin{aligned}
        &\sigma^2 &\quad (i \notin \Lambda_s)\\
        &\sigma^2 (1+\alpha) &\quad (i \in \Lambda_s)
    \end{aligned}
    \right.
.
\end{align}
% \begin{align}
% \label{equ:lambda_i}
% \lambda_i(\boldsymbol{x},s)
% &\propto \exp\left(
% -\frac{
% \left(x-\mu_{i,x}(s)\right)^2
% +\left(y-\mu_{i,y}(s)\right)^2
% }{2\sigma_i(s)^2}
% \right),\\
% \sigma_i(s)^2
% &= \left\{
% \begin{aligned}
% &\sigma^2 &\quad (i \notin \Lambda_s)\\
% &\sigma^2 (1+\alpha) &\quad (i \in \Lambda_s)
% \end{aligned}
% \right.
% \end{align}
Here, $\sigma$ is the baseline place-field width, and $\alpha$ quantifies the relative increase in place-field size for an engram cell in its associated context.
The vector $\boldsymbol{\mu}_i(s)$ denotes the center of the place field of the $i$th cell in context $C_s$.
% The vector $(\mu_{i,x}(s),\mu_{i,y}(s))^\top$ denotes the center of the place field of the $i$th cell in context $C_s$.
We assume that these place-field centers are independently and uniformly distributed within each environment.

\subsection{Mutual information}

To quantify the information encoded by the population spiking activity, we use mutual information\cite{shannon1948mathematical, cover1991elements}. Let $X$ denote the index of the firing neuron, which takes values from $1$ to $N$.
The spatial information carried by a single observed spike is quantified by mutual information, more specifically the conditional mutual information between $X$ and the animal's position $\boldsymbol{x}$ conditioned on the context $s$:
\begin{align}
  \label{equ:Ispace_def}
I_\mathrm{space} &= I(\boldsymbol{x}; X | s)\notag\\
&=
\sum_{s=0}^{M} P(s)
\int d\boldsymbol{x}\sum_{i=1}^{N}P(i|\boldsymbol{x},s)P(\boldsymbol{x})\log\frac{P(i|\boldsymbol{x},s)}{P(i|s)}.
\end{align}
% The spatial information carried by a single observed spike is quantified by the mutual information between $X$ and the animal's position $\boldsymbol{x}$:
% \begin{align}
%   \label{equ:Ispace_def}
% I_\mathrm{space} &= I(\boldsymbol{x}; X)\notag\\
% &= \int d\boldsymbol{x}\sum_{i=1}^{N}P(i|\boldsymbol{x})P(\boldsymbol{x})\log\frac{P(i|\boldsymbol{x})}{P(i)}.
% \end{align}
Similarly, the contextual information carried by a single observed spike is given by the conditional mutual information between $X$ and the context variable $s$ conditioned on the animal's position $\boldsymbol{x}$:
\begin{align}
  \label{equ:Icontext_def}
I_\mathrm{context} &= I(s; X|\boldsymbol{x})\notag\\
&=
\int d\boldsymbol{x} P(\boldsymbol{x})
\sum_{s=0}^{M}\sum_{i=1}^{N}P(i|s,\boldsymbol{x})P(s)\log\frac{P(i|s,\boldsymbol{x})}{P(i|\boldsymbol{x})}.
\end{align}
% Similarly, the contextual information carried by a single observed spike is given by the mutual information between $X$ and the context variable $s$:
% \begin{align}
%   \label{equ:Icontext_def}
% I_\mathrm{context} &= I(s; X)\notag\\
% &=\sum_{s=0}^{M}\sum_{i=1}^{N}P(i|s)P(s)\log\frac{P(i|s)}{P(i)}.
% \end{align}

Although $I(s;X|\boldsymbol{x})$ cannot be evaluated analytically, the statistical independence of $s$ and $\boldsymbol{x}$ in this setting allows us to show that the mutual information $I(s;X)$ provides a lower bound on $I(s;X|\boldsymbol{x})$. We therefore use this lower bound in place of $I(s;X|\boldsymbol{x})$ in the following analysis (see Supplemental Materials for details):
\begin{align}
I_\mathrm{context}
= I(s;X|\boldsymbol{x})
\ge I(s;X).
\label{equ:Icontext_bound}
\end{align}

Because the animal's position $\boldsymbol{x}$ is continuous whereas the context variable $s$ is discrete, $I_\mathrm{space}$ and $I_\mathrm{context}$ are not directly comparable. We therefore define the total information carried by a single observed spike as the weighted sum,
\begin{align}
I_\mathrm{total}
= W I_\mathrm{space} + I_\mathrm{context},
\label{equ:Itotal}
\end{align}
where $W$ specifies the relative importance assigned to spatial information compared with contextual information.

We then consider the increase in total information resulting from the recruitment of place cells as engram cells relative to a reference network in which no CA1 place cell is recruited as an engram cell, i.e., $f_s=0$ for all $s$. Because both $I(s;X|\boldsymbol{x})$ and $I(s;X)$ vanish in this reference network, the difference in $I(s;X)$ between a network with engram cell recruitment and the reference network indeed provides a lower bound on the increase in contextual information. Accordingly, we can use the following quantity as a lower bound on the increase in total information:
\begin{align}
\Delta I_\mathrm{total}^{\rm LB}
= W\Delta I_\mathrm{space}+\Delta I(s;X),
\label{equ:deltaItotal}
\end{align}
where $\Delta I_{\rm space}$ and $\Delta I(s;X)$ denote the differences in $I_\mathrm{space}$ and $I(s;X)$, respectively, between a network in which a fraction $f_s=|\Lambda_s|/N$ of place cells is recruited as engram cells for context $C_s$ and the reference network. In the following analysis, we use $\Delta I_\mathrm{total}^{\rm LB}$ to evaluate the increase in total information.

% Because the animal's position $\boldsymbol{x}$ is continuous whereas the context variable $s$ is discrete, $I_\mathrm{space}$ and $I_\mathrm{context}$ are not directly comparable. We therefore define the total information carried by a single observed spike as the weighted sum,
% \begin{align}
%   \label{equ:Itotal}
% I_\mathrm{total} = W I_\mathrm{space} + I_\mathrm{context}
% ,
% \end{align}
% where $W$ specifies the relative importance assigned to spatial information compared with contextual information. We then define the increase in total information resulting from the recruitment of place cells as engram cells as:
% \begin{align}
%   \label{equ:deltaItotal}
% \Delta I_\mathrm{total} = W \Delta I_\mathrm{space} + \Delta I_\mathrm{context}
% ,
% \end{align}
% where $\Delta I_{\rm space}$ and $\Delta I_{\rm context}$ denote the differences in spatial and contextual information, respectively, between a network in which a fraction $f_s=|\Lambda_s|/N$ of place cells is recruited as engram cells for context $C_s$ and a reference network in which no CA1 place cell is recruited as an engram cell, i.e., $f_s=0$ for all $s$.

Applying Bayes' theorem, we analytically derive an explicit expression for the lower bound on the increase in total information as a function of four sets of variables, $\Delta I_\mathrm{total}^{\rm LB}(\alpha,\{f_s\},\{p_s\},W)$ (see Supplemental Materials for details). These variables are (i) $\alpha$, which quantifies the relative increase in place-field size for engram cells and thus characterizes the strength of contextual input; (ii) $\{f_s\}$, the set of engram cell fractions for each context; (iii) $\{p_s\}$, the set of occurrence probabilities of the contexts; and (iv) $W$, the relative weight assigned to spatial information compared with contextual information, which characterizes the importance of spatial information for the animal.

Finally, the optimal allocation of engram cells, namely the set of engram cell fractions $\{f_s\}$ assigned to the individual contexts, is obtained by solving the following optimization problem:
\begin{align}
\max_{\{f_s\}} \quad &
\Delta I_\mathrm{total}^{\rm LB}(\alpha,\{f_s\},\{p_s\},W)\notag\\
\mathrm{s.t.}\quad
&0\le f_s\le1
\quad(s=1,\ldots,M),\label{equ:optimization}\\
&\sum_{s=1}^{M}f_s\le1.\notag
\end{align}
The first constraint follows from the definition of $f_s$ as the fraction of place cells recruited as engram cells for context $C_s$, whereas the second ensures that the sets of engram cells ${\Lambda_s}$ are mutually disjoint. Note that the optimal fraction $f_s$ generally depends on the entire set of context probabilities $\{p_s\}$ rather than only on the corresponding probability $p_s$. We therefore solve the optimization problem numerically for various choices of $\{p_s\}$ to investigate how the optimal engram cell fractions depend on the context probabilities.

\subsection{Analytical derivation of the phase boundary for finite engram cell allocation}

To analytically determine the boundary of the phase in which no engram cells are allocated to any context in the $(\alpha, W)$ plane shown in Figure~\ref{fig:4}, we apply the Karush--Kuhn--Tucker (KKT) conditions\cite{vandenberghe2004convex} to the optimization problem, Eq.~\ref{equ:optimization}. This boundary separates the phase with no engram cell allocation from the phase in which a finite fraction of place cells is recruited as engram cells for at least one context.

The KKT conditions for the optimization problem are given by
\begin{align}
\left\{
\begin{aligned}
    \frac{\partial \Delta I_\mathrm{total}^{\rm LB}}{\partial f_s}\Big|_{f_s=f_s^*} + \nu_s - \mu &= 0\\
    \nu_s \ge 0, \mu &\ge 0\\
    f_s^* \nu_s &= 0\\
    \left(\sum_{s=1}^{M}f_s^* - 1\right) \mu &= 0
\end{aligned}
\right.
\label{equ:KKT}
\end{align}
where $\nu_s$ $(s=1,\dots,M)$ and $\mu$ are the KKT multipliers. When the optimal solution satisfies $f_s^*=0$ for all contexts, the last condition requires $\mu=0$. Substituting $\mu=0$ into the first two conditions gives
\begin{align}
\frac{\partial \Delta I_\mathrm{total}^{\rm LB}}{\partial f_s}\Big|_{f_s=0}
&=
\left[(1-W)(1+\alpha)\log(1+\alpha)\right]p_s
-(1+\alpha p_s)\log(1+\alpha p_s)
\notag\\
&=-\nu_s\le 0
\label{equ:suffcondition}
\end{align}
for all $s=1,\dots,M$.

If these inequalities hold for all $p_s\in[0,1]$, the optimal engram cell fractions are guaranteed to satisfy $f_s^*=0$ for all contexts, irrespective of the context occurrence probabilities. Using the inequality
\begin{align}
\alpha \le
\frac{(1+\alpha p_s)\log(1+\alpha p_s)}{p_s},
\notag
\end{align}
which holds for all $p_s\in(0,1]$, with the equality satisfied in the limit $p_s\to0$, we obtain the following sufficient condition for the absence of engram cell allocation:
\begin{align}
W \ge
1-\frac{\alpha}{(1+\alpha)\log(1+\alpha)}.
\label{equ}
\end{align}
This condition provides the analytical boundary shown in Figure~\ref{fig:4}.

\section{Results}

\subsection{Optimal engram cell allocation for multiple contexts}

To analytically investigate the trade-off between spatial and contextual information associated with engram cell allocation, we first quantify both types of information using mutual information, more specifically conditional mutual information, and derive an explicit expression for a lower bound on the increase in total information encoded by population spiking activity due to engram cell allocation. We then numerically solve the optimization problem that maximizes this lower bound to obtain the optimal engram cell fractions $\{f_s\}$ as functions of the context occurrence probabilities $\{p_s\}$, the relative increase in place-field size for engram cells $\alpha$, and the relative importance assigned to spatial information $W$ (see Methods and Supplemental Materials for details).
% To analytically investigate the trade-off between spatial and contextual information associated with engram cell allocation, we first quantify both types of information using mutual information and derive an explicit expression for the increase in total information encoded by population spiking activity due to engram cell allocation.
% % To analytically investigate the trade-off between spatial and contextual information associated with engram cell allocation, we first derive an explicit expression for the increase in total information encoded by population spiking activity due to engram cell allocation. 
% We then numerically solve the optimization problem that maximizes the increase in total information to obtain the optimal engram cell fractions $\{f_s\}$ as functions of the context occurrence probabilities $\{p_s\}$, the relative increase in place-field size for engram cells $\alpha$, and the relative importance assigned to spatial information $W$ (see Methods and Supplemental Materials for details). 
Thus, $f_s$ denotes the optimal fraction of place cells recruited as engram cells for context $C_s$, which occurs with probability $p_s$.
% Here, $f_s$ denotes the optimal fraction of place cells recruited as engram cells for context $C_s$ with occurrence probability $p_s$.

Figure \ref{fig:2} shows the numerically obtained optimal engram cell fraction $f_s$ as a function of the context occurrence probability $p_s$. The results reveal three intriguing features of the optimal engram cell allocation. First, the optimal engram cell fraction decreases with increasing context occurrence probability when the occurrence probability is sufficiently large. This behavior can be understood as a consequence of the trade-off between spatial and contextual information. Because recruiting place cells as engram cells reduces the spatial information encoded by the hippocampal population, allocating a large fraction of engram cells to frequently experienced contexts substantially decreases the total information. This strategy is also biologically plausible because animals that repeatedly experience the same environment benefit more from preserving accurate spatial representations than from allocating additional neural resources to contextual information.

Second, unlike the behavior at large context occurrence probabilities, the optimal engram cell fraction also decreases with decreasing context occurrence probability when $p_s$ is sufficiently small. Consequently, the optimal engram cell fraction exhibits a non-monotonic dependence on the context occurrence probability, reaching its maximum at an intermediate probability.

This behavior cannot be explained by the trade-off between spatial and contextual information alone because recruiting place cells as engram cells for rarely experienced contexts only weakly degrades the spatial information encoded by the hippocampal population. Instead, it arises from competition among different contexts for the limited neural resources available for contextual representation. Because neural resources must be distributed across all contexts, allocating many engram cells to a rarely experienced context inevitably reduces the resources available for other contexts. Therefore, the optimal strategy is to allocate fewer engram cells to contexts that are encountered only infrequently.

To confirm this interpretation, we performed an additional analysis in which the competition among contexts was removed (see Appendix~\ref{appendix:single}). As expected, in the absence of inter-context competition, the optimal engram cell fraction became a monotonically decreasing function of the context occurrence probability.

Third, surprisingly, the optimal engram cell fraction undergoes a discontinuous transition from a finite value to zero at a critical context occurrence probability. Consequently, no engram cells are allocated to contexts whose occurrence probabilities are below the critical value. This behavior also arises from competition among different contexts and disappears when the competition is removed. This strategy is also biologically plausible because maintaining memories of environments that are extremely unlikely to be encountered again is an inefficient use of limited neural resources. Instead, the limited neural resources can be allocated to representing frequently experienced contexts and preserving accurate spatial representations.

\myfigure[width=8cm]{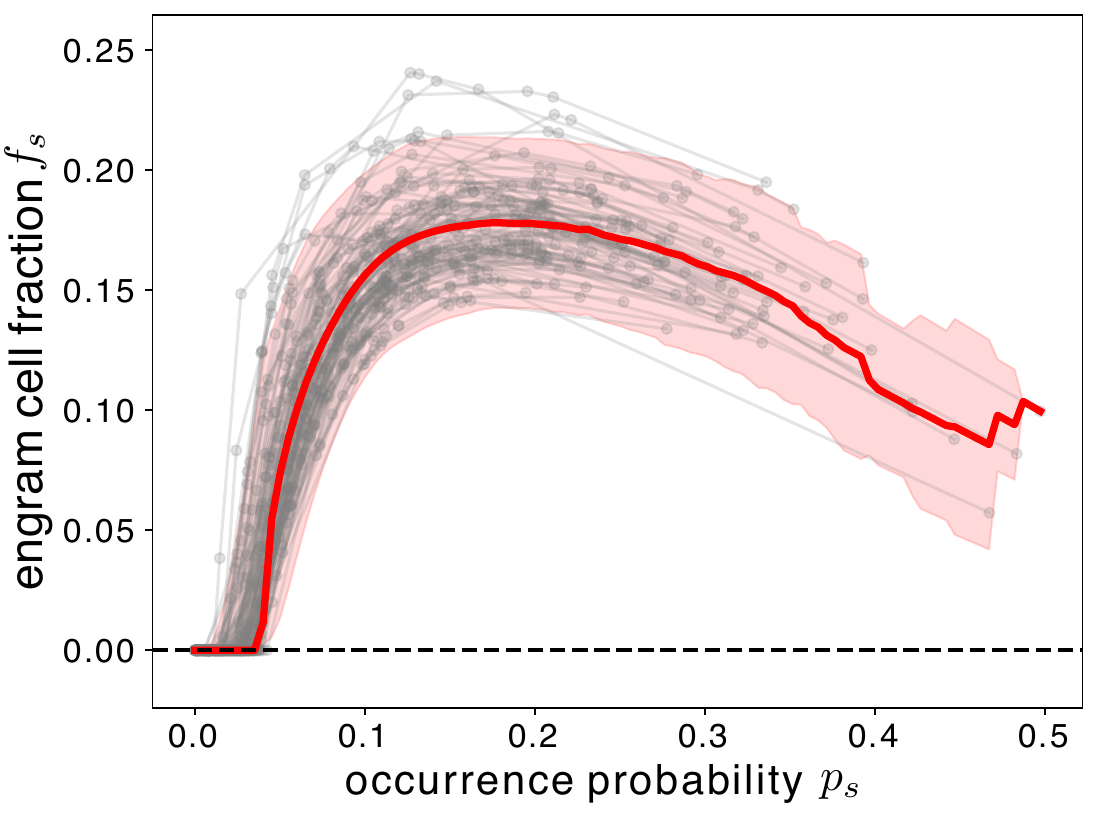}{Optimal engram cell fraction $f_s$ as a function of context occurrence probability $p_s$. Each set of gray points connected by a gray line represents the optimal solution of Eq.~\eqref{equ:optimization} for a specific realization of context occurrence probabilities. The red line represents the mean of a truncated normal distribution fitted by maximum likelihood, and the light red shaded region represents the 95\% confidence interval of the estimated mean. The result corresponding to the home cage ($s=0$) is excluded from the figure. Other parameters are $\alpha=0.5$, $W=0.1$, and $M=10$.}{2}

\subsection{Phase boundary between phases with and without engram cell allocation}

To investigate how engram cell allocation depends on the relative increase in place-field size for engram cells, $\alpha$, and the relative importance assigned to spatial information, $W$, we numerically solve the optimization problem for various combinations of these parameters and obtain the optimal engram cell fraction $f_s$ as a function of the context occurrence probability $p_s$.

Figure \ref{fig:3} shows the results.
As in Figure \ref{fig:2}, each panel shows the numerically obtained optimal engram cell fraction $f_s$ as a function of the context occurrence probability $p_s$, with different panels corresponding to different combinations of $\alpha$ and $W$.
For relatively small values of $W$, the optimal engram cell fraction remains finite and exhibits the same non-monotonic dependence on context occurrence probability as described in the previous subsection. As $W$ increases, however, the optimal solution undergoes a qualitative transition to a phase in which no engram cells are allocated for any value of $p_s$. Because $W$ represents the relative importance assigned to spatial information compared with contextual information, this result indicates that suppressing engram cell allocation is the information-theoretically optimal strategy when accurate spatial representation becomes sufficiently important for the animal.

The results further suggest that the critical value of $W$ increases with increasing $\alpha$. Because the relative increase in place-field size is expected to be induced by contextual input, $\alpha$ may be regarded as a measure of the strength of contextual input to hippocampal CA1 neurons. These observations suggest that stronger contextual input facilitates the emergence of engram cells, allowing finite engram cell allocation even when spatial information is assigned a relatively large weight.

The above observations suggest the existence of two distinct phases in the optimal allocation of engram cells. In one phase, no engram cells are allocated to any context, whereas in the other, a finite fraction of place cells is recruited as engram cells. To analytically derive the phase boundary between these phases, we apply the Karush-Kuhn-Tucker (KKT) conditions, which characterize the solution of a constrained optimization problem. Combining the KKT conditions with the condition $f_s=0$ for all $s$, we obtain the following inequality, which provides a sufficient condition on the parameters $W$ and $\alpha$ for the absence of engram cell allocation irrespective of the context occurrence probabilities (see Methods for details):
\begin{align}
W \ge 1 - \frac{\alpha}{(1+\alpha)\log(1+\alpha)} .\label{equ:transition}
\end{align}

To examine the analytical prediction, we numerically solved the optimization problem over a range of parameter values $(\alpha, W)$ and measured the maximum engram cell fraction, $\displaystyle f_\mathrm{max} = \max_s f_s$. The maximum engram cell fraction is positive when a finite fraction of engram cells is allocated to at least one context, whereas it is zero when no engram cells are allocated.

Figure~\ref{fig:4} shows the resulting phase diagram in the $(\alpha, W)$ plane, where the color indicates $\log_{10}(f_{\max})$. As predicted, $f_{\max}$ is finite for sufficiently small values of $W$ and becomes zero when $W$ is large. The critical value of $W$ increases with increasing $\alpha$. The thick red curve represents the analytically derived phase boundary given by Eq.~\eqref{equ:transition}. The analytical prediction accurately captures the numerically obtained phase boundary, showing good agreement between the theoretical prediction and the numerical results.

\myfigure[width=12cm]{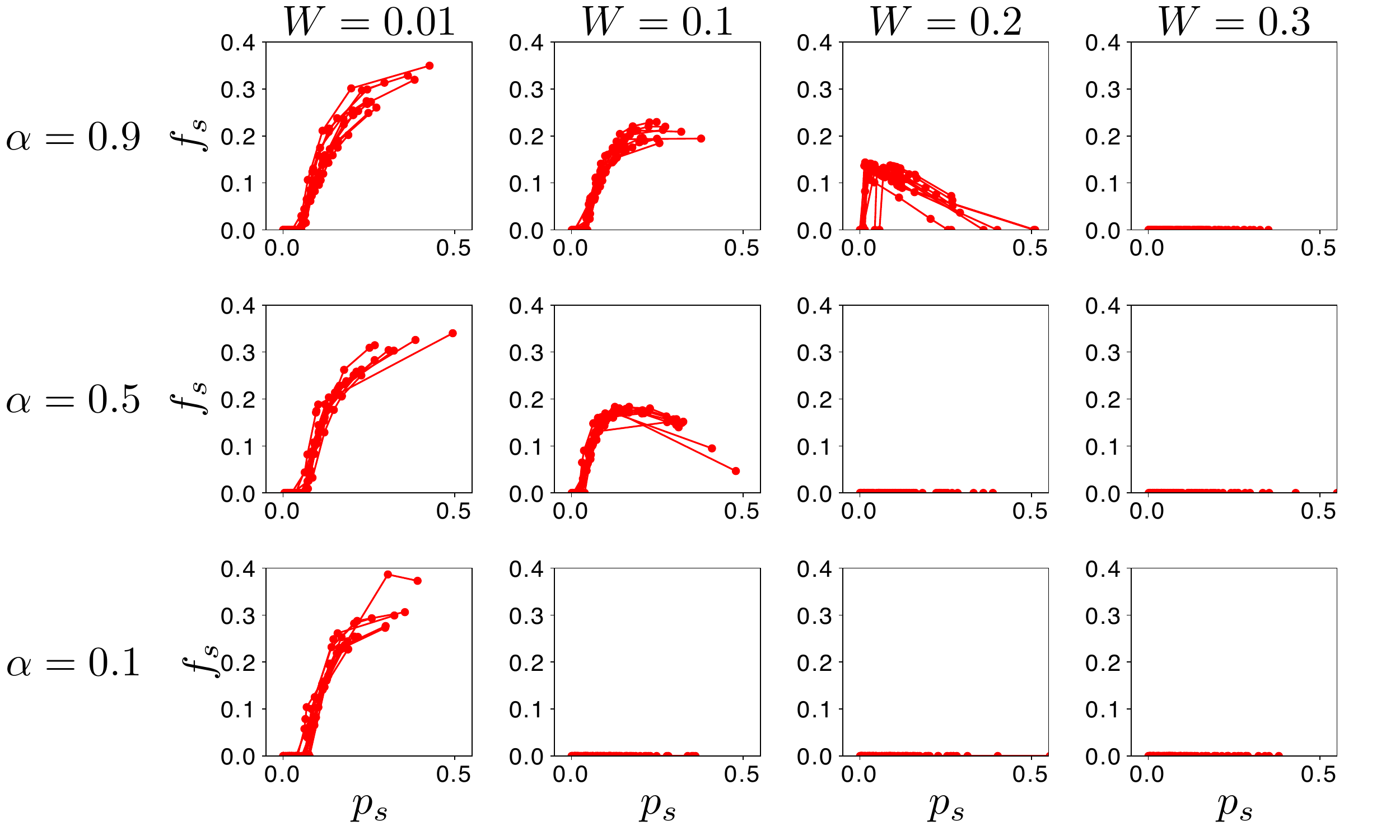}{Optimal engram cell fraction $f_s$ as a function of context occurrence probability $p_s$ for different values of the relative increase in place-field size, $\alpha$, and the relative importance assigned to spatial information, $W$. Each panel corresponds to different values of $\alpha$ and $W$. As in Fig.~2, the red points connected by a red line in each panel show the optimal solution $f_s$ as a function of $p_s$ for a specific realization of context occurrence probabilities ${p_s}$. As $W$ increases from the left panels to the right panels, the optimal solution undergoes a qualitative transition to a state in which no engram cells are allocated for any context. The critical value of $W$ at which this transition occurs increases with increasing $\alpha$, from the bottom panels to the top panels.}{3}

\myfigure[width=14cm]{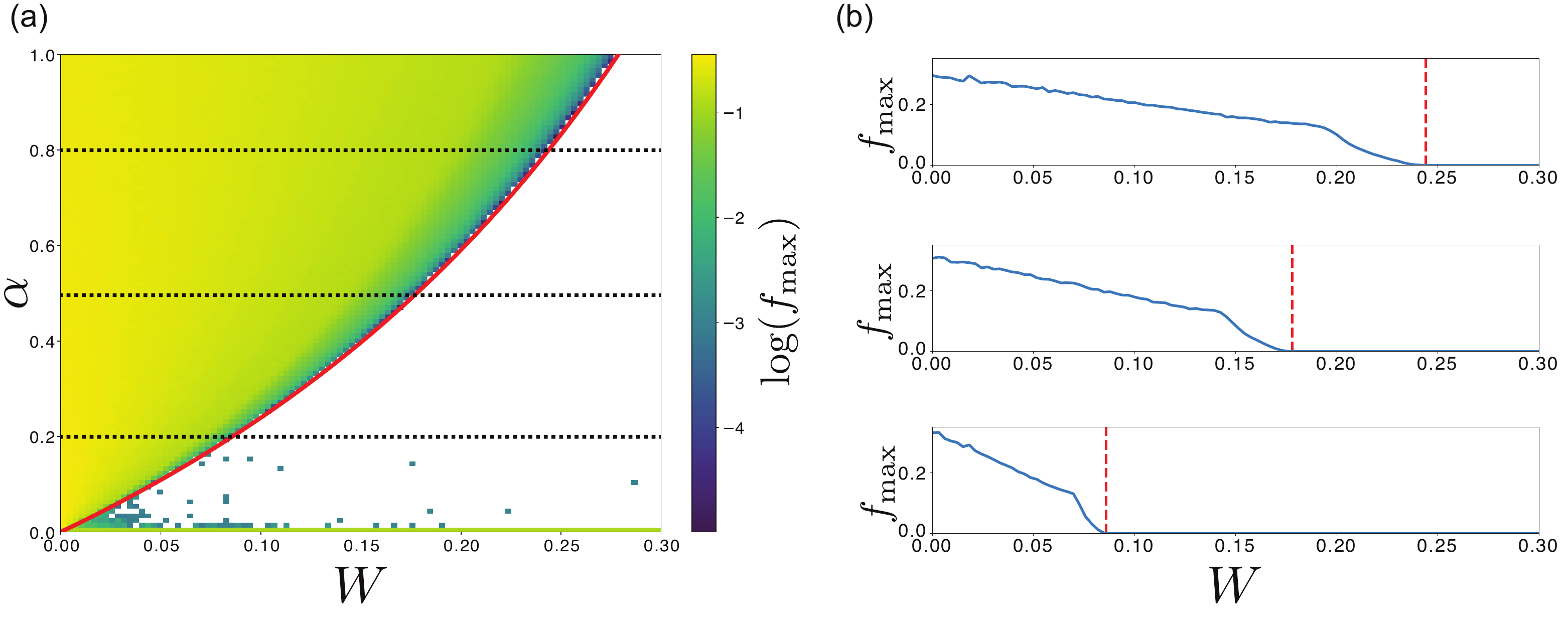}{Phase transition of the recruited engram cell fraction. (a) Phase diagram of the numerically obtained maximum optimal engram cell fraction $f_{\max}$ over various realizations of the context occurrence probabilities ${p_s}$. The maximum fraction $f_{\max}$ undergoes a qualitative transition and vanishes in the region with larger $W$ and smaller $\alpha$. The color of the heatmap represents $\log_{10}(f_{\max})$. Regions where $f_{\max}<10^{-5}$ are shown in white because $\log_{10}(f_{\max})$ diverges to $-\infty$ as $f_{\max}$ approaches zero. The thick red line represents the theoretical prediction of the phase boundary given by Eq.~\eqref{equ:transition}. The horizontal dotted lines indicate the values of $\alpha$ used in panel (b). (b) $f_{\max}$ as a function of $W$ for $\alpha=0.8$ (top), $\alpha=0.5$ (middle), and $\alpha=0.2$ (bottom). The red vertical line in each panel indicates the theoretically derived transition point.}{4}
\section{Discussion}
In this study, we developed an information-theoretic framework to investigate how CA1 neurons should be allocated to represent spatial and contextual information,. Based on experimental observations that hippocampal place cells are recruited as engram cells in CA1 accompanied by an expansion of their place fields\cite{engram_PC}, we formulated the trade-off between spatial and contextual information using the mutual information between CA1 spiking activity and the animal's spatial location or contextual experience.

By optimizing the total information carried by CA1 place cells, we found that the optimal fraction of place cells recruited as engram cells depends non-monotonically on the context occurrence probability. The optimal fraction undergoes a discontinuous transition from zero to a finite value at a critical probability, reaches a maximum at intermediate probabilities, and decreases for highly frequent contexts. Furthermore, using the Karush--Kuhn--Tucker (KKT) conditions, we identified two distinct phases depending on the relative increase in place-field size and the relative importance assigned to spatial information. In one phase, no engram cells are allocated to any context, whereas in the other, a finite fraction of place cells is recruited as engram cells for at least one context.

The decrease in the optimal engram cell fraction for contexts with high occurrence probabilities may be interpreted in relation to memory consolidation in the brain\cite{engram_review2, kitamura2017engrams, klinzing2019mechanisms, brodt2023sleep}. The hippocampus plays a central role in the representation of episodic memories\cite{moscovitch2016episodic}, particularly during their initial formation and retrieval\cite{nakazawa2003hippocampal, frankland2005organization}, whereas memories can become increasingly dependent on neocortical representations over longer time scales through systems consolidation. Contexts that are encountered frequently are likely to accumulate repeated experiences over time, which may facilitate their consolidation into neocortical representations. Thus, memories of frequently experienced contexts may become less dependent on hippocampal representations as they are consolidated in the neocortex. This possibility is consistent with our finding that the optimal fraction of hippocampal engram cells decreases as the context occurrence probability increases.

In the present study, we developed our theory under a stationary setting with fixed occurrence probabilities for each contextual event. The optimal engram cell fractions obtained in this framework can therefore be interpreted as those acquired in CA1 after the animal has sufficiently experienced the different contexts at given frequencies. However, animals do not necessarily experience contexts under stationary conditions, and the effective occurrence probabilities of individual contexts may continuously change with experience. Extending the present framework to such a dynamic setting may provide a possible interpretation of memory forgetting\cite{ryan2022forgetting, richards2017persistence}. Our results show that the optimal engram cell fraction decreases as the occurrence probability becomes small and eventually vanishes below a critical probability. This could imply that, as the time since the last experience of a context increases and its effective occurrence probability decreases, its representation by engram cells in CA1 gradually diminishes and eventually disappears. Thus, the disappearance of engram cell allocation predicted by our theory may provide an information-theoretic description of memory forgetting. Formulating the optimal allocation of neurons to memories in such a dynamic setting, including the processes of memory formation and forgetting, will be an important future direction.

To our knowledge, the relationship between context occurrence probability and the fraction of engram cells has not yet been experimentally investigated under the stationary conditions considered here. It would therefore be interesting to determine whether the non-monotonic dependence predicted by our theory can be observed experimentally. On the other hand, an experimental study has examined how the number of engram cells changes with the duration of context exposure in a setting closer to the dynamic scenario discussed above\cite{leake2021engram}. This study reported that the number of engram cells increases with exposure duration. This observation may be consistent with our finding that the optimal engram cell fraction increases with context occurrence probability up to intermediate probabilities, although our theory does not account for the process of engram cell recruitment during learning. Extending the information-theoretic framework to describe the temporal evolution of engram cell allocation during learning will be an important direction for future work.

Another prediction of our theory is that engram cell recruitment decreases as the relative increase in place-field size $\alpha$ decreases or the relative importance assigned to spatial information $W$ increases, and eventually disappears beyond the phase boundary. Direct manipulation of $\alpha$ may be difficult. (If $\alpha$ reflects the strength of contextual input to hippocampal neurons, it could potentially be varied by modulating such input.) In contrast, manipulating the relative importance of spatial information may be more experimentally feasible. For example, behavioral tasks could be designed in which accurate identification of spatial location is required to obtain a reward, thereby increasing the importance of spatial information relative to contextual information. Such experiments may provide a way to examine the predicted decrease and eventual disappearance of engram cell recruitment under different behavioral conditions.

Our theory also relies on several simplifying assumptions introduced for analytical tractability. We assumed that engram cells representing different contexts are mutually disjoint, although individual neurons may represent multiple contexts in practice. We also assumed fixed place-field sizes for place cells and engram cells, whereas place-field sizes are likely to be heterogeneous across neurons. (More generally, place cells and engram cells may not necessarily constitute two distinct populations, but instead may form a continuum in their physiological properties, including place-field size.) 
In addition, our analysis is based on a lower bound on the increase in total information, and it remains unclear how close this bound is to the actual information increase. Extending the present framework to more realistic conditions and evaluating the tightness of this bound remain important challenges.
% Extending the present framework to more realistic conditions remains an important challenge.

A key feature of our framework is the use of mutual information to formulate efficient memory representations. While information-theoretic approaches have been widely used to study efficient coding in sensory systems, their implications for memory representations have been much less explored. The present study extends this perspective to memory by asking how limited neural resources can be efficiently allocated to encode spatial and contextual information. However, whether this framework can be extended to memory-guided behavior remains an open question. In particular, it remains unclear whether a memory representation that is optimal in terms of mutual information also supports efficient retrieval and appropriate behavior based on the retrieved memory. Exploring the information-theoretic link between memory representation and memory-guided behavior will be an important future direction for both theoretical and experimental neuroscience.

\bibliographystyle{unsrt}
\bibliography{myrefs}

\appendix
% 図の番号フォーマットを「S1, S2, ...」に変更
\renewcommand{\thefigure}{S\arabic{figure}}
% 図のカウンターを1（S1）から始まるようにリセット
\setcounter{figure}{0}
\section{Derivation of the analytic expressions for mutual information}
\label{appendix:info}

Let $\lambda_i^+(\boldsymbol{x})$ and $\lambda_i^-(\boldsymbol{x})$ denote the place fields, i.e., the firing rates, of the $i$th neuron depending on whether it is recruited as an engram cell in a given context:
\begin{align*}
\lambda_i(\boldsymbol{x},s)
&=
\left\{
    \begin{aligned}
    A\exp\left(
        -\frac{
        \left|\boldsymbol{x}-\boldsymbol{\mu}_i(s)\right|^2
        }{2\sigma^2(1+\alpha)}
    \right)
    &=: \lambda_i^+(\boldsymbol{x}) \quad (i\in\Lambda_s)
    \\
    A\exp\left(
        -\frac{
        \left|\boldsymbol{x}-\boldsymbol{\mu}_i(s)\right|^2
        }{2\sigma^2}
    \right)
    &=: \lambda_i^-(\boldsymbol{x}) \quad (i\notin\Lambda_s)
    \end{aligned}
\right.
.
\end{align*}
Because the conditional probability that an observed spike is generated by the $i$th neuron is proportional to its firing rate, we obtain
\begin{align}
P(i|\boldsymbol{x},s)
=
\frac{\lambda_i(\boldsymbol{x},s)}
{\sum_{j=1}^{N}\lambda_j(\boldsymbol{x},s)}
=
\left\{
\begin{aligned}
    \frac{\lambda_i^+(\boldsymbol{x})}{r_s}
    &\quad (i\in\Lambda_s),\\
    \frac{\lambda_i^-(\boldsymbol{x})}{r_s}
    &\quad (i\notin\Lambda_s)
\end{aligned}
\right.
,
\label{equ:P(i|x)}
\end{align}
where, assuming that the number of neurons in CA1 is sufficiently large, we replace the sum of their firing rates by its ensemble average,
$r_s:=N\left(f_s\lambda^+ +(1-f_s)\lambda^-\right)$.

We next define the firing rate marginalized over the spatial position $\boldsymbol{x}$ as $\lambda_i(s)$. Defining $\lambda^+$ and $\lambda^-$ accordingly, we obtain
\begin{align}
\lambda_i(s)
&=
\int d\boldsymbol{x}\,
\lambda_i(\boldsymbol{x},s)P(\boldsymbol{x}|s)
\notag\\
&=
\left\{
\begin{aligned}
\int d\boldsymbol{x}\,
\frac{\lambda_i^+(\boldsymbol{x})}{S}
&=
\frac{2\pi\sigma^2(1+\alpha)A}{S}
=: \lambda^+
&\quad (i\in\Lambda_s)
\\
\int d\boldsymbol{x}\,
\frac{\lambda_i^-(\boldsymbol{x})}{S}
&=
\frac{2\pi\sigma^2A}{S}
=: \lambda^-
&\quad (i\notin\Lambda_s)
\end{aligned}
\right.
,
\label{equ:lambda}
\end{align}
where we used $P(\boldsymbol{x}|s)=P(\boldsymbol{x})=1/S$. Using Eqs.~\eqref{equ:P(i|x)} and \eqref{equ:lambda}, we obtain the marginalized conditional probability as
\begin{align}
P(i|s)
=
\int d\boldsymbol{x}\,
P(i|\boldsymbol{x},s)P(\boldsymbol{x}|s)
=
\left\{
\begin{aligned}
\frac{\lambda^+}{r_s}
=
\frac{1+\alpha}{N(1+\alpha f_s)}
&\quad (i\in\Lambda_s)\\
\frac{\lambda^-}{r_s}
=
\frac{1}{N(1+\alpha f_s)}
&\quad (i\notin\Lambda_s)
\end{aligned}
\right.
.
\label{equ:P(i)space}
\end{align}

Combining Eqs.~\eqref{equ:P(i|x)}, \eqref{equ:lambda}, and \eqref{equ:P(i)space} yields the mutual information between the spiking neuron index $X$ and the animal's position $\boldsymbol{x}$ conditioned on the context $s$:
\begin{align*}
I_\mathrm{space}
&=
\sum_{s=0}^{M}P(s)
\int d\boldsymbol{x}
\sum_{i=1}^{N}
P(i|\boldsymbol{x},s)P(\boldsymbol{x})
\log\frac{P(i|\boldsymbol{x},s)}{P(i|s)}
\\
&=
\sum_{s=0}^{M}
\frac{p_s}{r_sS}
\int d\boldsymbol{x}
\left[
\sum_{i\in\Lambda_s}
\lambda_i^+(\boldsymbol{x})
\log\frac{\lambda_i^+(\boldsymbol{x})}{\lambda^+}
+
\sum_{i\notin\Lambda_s}
\lambda_i^-(\boldsymbol{x})
\log\frac{\lambda_i^-(\boldsymbol{x})}{\lambda^-}
\right]
\\
&=
\sum_{s=0}^{M}
\frac{p_sNA}{r_sS}
\int d\boldsymbol{x}
\left[
f_s
e^{-\frac{|\boldsymbol{x}|^2}{2\sigma^2(1+\alpha)}}
\left(
\log\frac{A}{\lambda^+}
-\frac{|\boldsymbol{x}|^2}{2\sigma^2(1+\alpha)}
\right)
\right.
\\
&\hspace{42mm}\left.
+
(1-f_s)
e^{-\frac{|\boldsymbol{x}|^2}{2\sigma^2}}
\left(
\log\frac{A}{\lambda^-}
-\frac{|\boldsymbol{x}|^2}{2\sigma^2}
\right)
\right]
\\
&=
\left(
\log\frac{S}{2\pi\sigma^2}-1
\right)
-
\sum_{s=0}^{M}
p_s
\frac{(1+\alpha)f_s}{1+\alpha f_s}
\log(1+\alpha),
\end{align*}
where $f_s=|\Lambda_s|/N$. Because the first term is independent of $f_s$, the difference in spatial information between a network in which a fraction $f_s$ of place cells is recruited as engram cells and the reference network in which no engram cells are recruited is given by
\begin{align*}
\Delta I_\mathrm{space}
=
-\sum_{s=0}^{M}
p_s
\frac{(1+\alpha)f_s}{1+\alpha f_s}
\log(1+\alpha).
\end{align*}

Next, we turn to the contextual information carried by the observed spiking neuron index:
\begin{align}
I_\mathrm{context}
&=
I(s;X|\boldsymbol{x})
\notag\\
&=
\int d\boldsymbol{x}\,P(\boldsymbol{x})
\sum_{s=0}^{M}\sum_{i=1}^{N}
P(i|s,\boldsymbol{x})P(s)
\log\frac{P(i|s,\boldsymbol{x})}{P(i|\boldsymbol{x})}.
\end{align}
This conditional mutual information is not analytically tractable because
$\log P(i|\boldsymbol{x})=\log\sum_{s=0}^{M}p_sP(i|s,\boldsymbol{x})$
is the logarithm of a mixture of Gaussians, which cannot be integrated analytically. However, owing to the statistical independence between $s$ and $\boldsymbol{x}$, it can be shown that the unconditioned mutual information $I(s;X)$ provides a lower bound on the conditional mutual information:
\begin{align*}
I(s;X|\boldsymbol{x})
&=
H(s|\boldsymbol{x}) - H(s|X,\boldsymbol{x})
\\
&=
H(s) - H(s|X,\boldsymbol{x})
\\
&\ge
H(s) - H(s|X)
\\
&=
I(s;X)
\\
&=
\sum_{s=0}^{M}\sum_{i=1}^{N}
P(i|s)P(s)
\log\frac{P(i|s)}{P(i)},
\end{align*}
where $H(\cdot)$ and $H(\cdot|\cdot)$ denote entropy and conditional entropy, respectively. We used the independence between $s$ and $\boldsymbol{x}$ in the second line and the non-negativity of the conditional mutual information,
\begin{align*}
H(s|X) - H(s|X,\boldsymbol{x})
= I(s;\boldsymbol{x}|X)
\ge 0,
\end{align*}
in the third line. Thus, the lower bound is tight when $s$ and $\boldsymbol{x}$ are conditionally independent given $X$, with equality holding in this case.

To evaluate $I(s;X)$, we decompose it into two terms:
\begin{align}
I(s;X)
&=
\sum_{s=0}^{M} p_s \sum_{i=1}^{N} P(i|s) \log P(i|s)
-
\sum_{i=1}^{N} P(i) \log P(i).
\label{equ:I_context_multidef}
\end{align}
Using Eq.~\eqref{equ:P(i)space}, the first term is given by
\begin{align}
&\sum_{s=0}^{M}p_s \sum_{i=1}^{N}P(i|s)\log P(i|s)\notag\\
&=
\sum_{s=0}^{M}p_s
\left\{
\sum_{i\in\Lambda_s}P(i|s)\log P(i|s)
+
\sum_{i\notin\Lambda_s}P(i|s)\log P(i|s)
\right\}\notag\\
&=
\sum_{s=0}^{M}p_s
\left\{
f_s N \frac{1+\alpha}{N(1+\alpha f_s)}
\log\frac{1+\alpha}{N(1+\alpha f_s)}
+
(1-f_s)N \frac{1}{N(1+\alpha f_s)}
\log\frac{1}{N(1+\alpha f_s)}
\right\}\notag\\
&=
\sum_{s=0}^{M}p_s
\left\{
\frac{(1+\alpha)f_s}{1+\alpha f_s}\log(1+\alpha)
-\log(1+\alpha f_s)
\right\}
-\log N.
\label{equ:I_context_multi_first}
\end{align}

Eq.~\eqref{equ:P(i)space} also gives the marginal probability
\begin{align}
P(i)
&=
\sum_{s=0}^{M}p_sP(i|s)\notag\\
&=
\sum_{s;\,i\in\Lambda_s}
\frac{(1+\alpha)p_s}{N(1+\alpha f_s)}
+
\sum_{s;\,i\notin\Lambda_s}
\frac{p_s}{N(1+\alpha f_s)},
\end{align}
where the first and second sums run over contexts in which the $i$th neuron is and is not recruited as an engram cell, respectively. Using this expression, the second term in Eq.~\eqref{equ:I_context_multidef} is calculated as
\begin{align}
&\sum_{i=1}^{N}P(i)\log P(i)\notag\\
&=
\sum_{s=0}^{M}\sum_{i\in\Lambda_s}P(i)\log P(i)
+
\sum_{i\notin\cup_{s=0}^{M}\Lambda_s}P(i)\log P(i)\notag\\
&=
\sum_{s=0}^{M}f_s N
\left\{
\frac{(1+\alpha)p_s}{N(1+\alpha f_s)}
+
\sum_{s'\ne s}\frac{p_{s'}}{N(1+\alpha f_{s'})}
\right\}\notag\\
&\quad\times
\log\left(
\frac{(1+\alpha)p_s}{N(1+\alpha f_s)}
+
\sum_{s'\ne s}\frac{p_{s'}}{N(1+\alpha f_{s'})}
\right)\notag\\
&\quad+
\left(1-\sum_{s=0}^{M}f_s\right)N
\left(
\sum_{s=0}^{M}\frac{p_s}{N(1+\alpha f_s)}
\right)
\log\left(
\sum_{s=0}^{M}\frac{p_s}{N(1+\alpha f_s)}
\right)\notag\\
&=
\sum_{s=0}^{M}f_s
\left\{
\frac{(1+\alpha)p_s}{1+\alpha f_s}
+
\sum_{s'\ne s}\frac{p_{s'}}{1+\alpha f_{s'}}
\right\}
\log\left(
\frac{(1+\alpha)p_s}{1+\alpha f_s}
+
\sum_{s'\ne s}\frac{p_{s'}}{1+\alpha f_{s'}}
\right)\notag\\
&\quad+
\left(1-\sum_{s=0}^{M}f_s\right)
\left(
\sum_{s=0}^{M}\frac{p_s}{1+\alpha f_s}
\right)
\log\left(
\sum_{s=0}^{M}\frac{p_s}{1+\alpha f_s}
\right)
-\log N.
\label{equ:I_context_multi_second}
\end{align}
In the second line, we divide the neurons into those recruited as engram cells and those not recruited in any context. Because the engram-cell sets ${\Lambda_s}$ are mutually disjoint, each recruited neuron belongs to only one $\Lambda_s$ and is therefore counted only once in the first term. For a neuron belonging to $\Lambda_s$, the same assumption implies that it does not belong to $\Lambda_{s'}$ for any $s'\ne s$, yielding the sum over $s'\ne s$ in the third line. For a neuron that is not recruited as an engram cell for any context, the sum contains only the non-engram contribution from all contexts.

Substituting Eqs.~\eqref{equ:I_context_multi_first} and \eqref{equ:I_context_multi_second} into Eq.~\eqref{equ:I_context_multidef}, we obtain
\begin{align*}
I(s; X) &= \sum_{s=0}^{M} p_s
\left\{
\frac{(1+\alpha)f_s}{1+\alpha f_s}\log(1+\alpha)
-\log(1+\alpha f_s)
\right\}
\\
&-
\left[
\sum_{s=0}^{M}f_s
    \left\{
    \frac{(1+\alpha)p_s}{1+\alpha f_s}
    +\sum_{s'\ne s}\frac{p_{s'}}{1+\alpha f_{s'}}
    \right\}
\right.
\log
\left(
    \frac{(1+\alpha)p_s}{1+\alpha f_s}
    +\sum_{s'\ne s}\frac{p_{s'}}{1+\alpha f_{s'}}
\right)
\\
&\qquad+
\left.
    \left(1-\sum_{s=0}^{M}f_s\right)
    \left\{
        \sum_{s=0}^{M}\frac{p_s}{1+\alpha f_s}
    \right\}\log
    \left\{
        \sum_{s=0}^{M}\frac{p_s}{1+\alpha f_s}
    \right\}
\right]
\end{align*}
Because $I(s;X)=0$ when $f_s=0$ for all $s$, it immediately follows that
\begin{align*}
\Delta I(s;X)=I(s;X).
\end{align*}

\section{Optimal engram cell allocation in the absence of competition among contexts}\label{appendix:single}

For sufficiently small context occurrence probability $p_s$, the optimal engram cell fraction $f_s$ decreases as $p_s$ decreases. To examine whether this behavior results from competition among different contexts, we consider the case with only one context, $s=1$, in addition to the home cage, $s=0$. Because no engram cells are recruited in the home cage, there is no competition between contexts for neural resources in this case.

Figure~\ref{fig:5} shows the optimal engram cell fraction $f_1$ as a function of the context occurrence probability $p_1$ for different values of $\alpha$ and $W$. Here, $p_0=1-p_1$. In contrast to the non-monotonic dependence observed in the presence of multiple contexts, most curves are monotonically decreasing with $p_1$. In particular, none of the curves exhibits the rapid decrease in $f_1$ as $p_1$ decreases at small $p_1$ observed in the multi-context case. These results indicate that the decrease in the optimal engram cell fraction at small context occurrence probabilities results from competition among different contexts.

% \section{Results of the case without competition}\label{appendix:single}
% If we consider the simplest case in which only one context other than the home cage is to be stored, i.e., $s\in\{0,1\}$, there is no competition among the contexts for recruiting the engram cells. 
% In this case, the parameters of the model reduce to four variables: the strength of contextual input $\alpha$, the fraction of engram cells assigned to the context $f = f_1$, the occurrence probability of the context $p = p_1$, and the relative weight of spatial information $W$.\par
% For fixed values of $\alpha$ and $W$, we numerically solved the optimization problem defined in Eq. \eqref{equ:optimization} while varying the context occurrence probability $p$.
% The resulting optimal fraction of engram cells $f$ is shown in Fig. \ref{fig:single-context}.
% The results show that the optimal engram fraction $f$ tends to decrease with context occurrence probability $p$, which is seen in the sufficiently large range of context probability in Fig.~\ref{fig:multi-context}. 
% However, different from the case with multiple contexts, Fig.~\ref{fig:single-context} shows that the optimal engram cell fraction has a peak in an extremely rarely experienced context.
% This is unreasonable as the memory of the rarely experienced context should be forgotten in most cases, suggesting that the competition among the contexts in the engram cell allocation plays an essential role in the reasonable prediction of our model.

\myfigure[width=12cm]{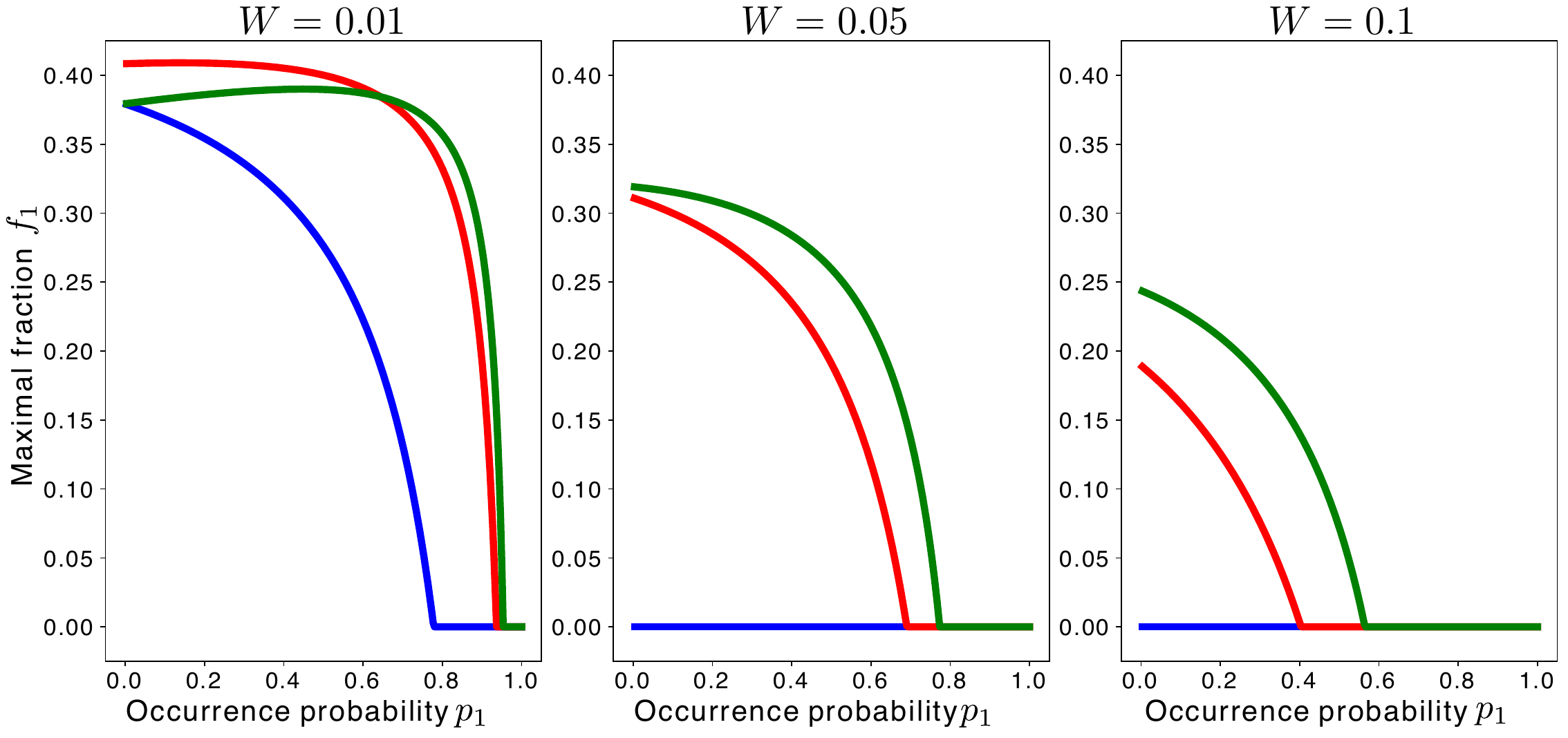}{Optimal engram cell fraction $f_1$ given by Eq.~\eqref{equ:optimization} as a function of context occurrence probability $p_1$ in the case with only one context, $s=1$, in addition to the home cage, $s=0$. Each panel shows the result for a different value of $W$, with $W=0.01$, $0.05$, and $0.1$ from left to right. The lines in each panel show the results for different values of $\alpha$: $\alpha=0.1$ (blue), $0.5$ (red), and $0.9$ (green).}{5}

% \section{The universality of the relationship between the engram cell fraction and the occurrence probability of the contexts}\label{appendix:universality}
% The optimal condition of the optimization problem \eqref{equ:optimization} is given by KKT condition \eqref{equ:KKT}. For a context $s$ with positive engram cell fraction $f_s$, the KKT condition becomes the equation below.
% \begin{align}
%     \frac{\partial \Delta I_\mathrm{total}}{\partial f_s}\Big|_{f_s=f^*_s} = \mu
% \end{align}
% $\frac{\partial \Delta I_\mathrm{total}}{\partial f_s}\Big|_{f_s=f^*_s}$ depends on the engram cell fraction and the occurrence probability of different conditions.
% As the dependence is only through the common factor $C=\sum_s\frac{p_s}{1+\alpha f_s}$ and $\mu$ is also common for different contexts, the relationship between the optimal engram cell fraction and the occurrence probability of the contexts is governed by the same equation:
% \begin{align}
%        \frac{\partial \Delta I_\mathrm{total}}{\partial f}\Big|_{f=f^*}(\alpha, f, p, W,C) = \mu(\alpha,f,p,W,C) 
% \end{align}
% Therefore, the approximate shape of the relationship between $f$ and $p$ is determined by $\frac{\partial \Delta I_\mathrm{total}}{\partial f}\Big|_{f=f^*}$, which leads to the universality of the shape throughout the different realization of $\{p_s\}$.

\end{document}